\PassOptionsToPackage{table}{xcolor}
\documentclass[submission,copyright,creativecommons]{eptcs}
\providecommand{\event}{FMAS 2026} 

\usepackage{iftex}
\usepackage{fretish}
\usepackage{graphicx}
\newcommand{\category}[1]{\textsc{#1}}

\usepackage{todonotes}

\usepackage{paralist}
\usepackage{xcolor}
\ifpdf
  \usepackage{underscore}         
  \usepackage[T1]{fontenc}        
\else
  \usepackage{breakurl}           
\fi

\title{Plenty to FRET About: From Hazard Analysis to Formal Requirements for Robot-Assisted Dressing}
\author{Gunjari Chakraborty \quad Yasmin Rafiq \quad Hazel M. Taylor \quad Marie Farrell
\institute{University of Manchester, UK}
}

\def\titlerunning{Plenty to FRET About: From Hazard Analysis to Formal Requirements for Robot-Assisted Dressing}
\def\authorrunning{G. Chakraborty, Y. Rafiq, H. M. Taylor \& M. Farrell}
\begin{document}
\maketitle

\begin{abstract}
Understanding requirements for systems involving humans and robots is essential for providing robust evidence that such systems behave correctly. Requirements formalisation approaches such as the Formal Requirements Elicitation Tool (FRET) can support the transition from unstructured natural language to formal representations suitable for verification. In this paper, we formalise 146 hazard-derived natural-language requirements for a Robot-Assisted Dressing (RAD) case study, previously classified using a taxonomy of Human-Robot Teamwork (HRT) requirements. Through an iterative process involving domain-expert clarification, these were represented by 117 unique \fretish{} requirements. Formalisation exposed ambiguities, temporal and probabilistic assumptions, and cases where source requirements needed to be decomposed or could be consolidated because they expressed the same formal behaviour. Temporal analysis showed that 70.1\% of the unique \fretish{} requirements use either immediate-response or ordering constraints (\timing{immediately} or \timing{before}), while formalisation also exposed relationships between requirements across HRT taxonomy categories. We further analyse the formal semantics and realisability of the resulting requirements. The results show that HRT classification and formalisation provide complementary views: the taxonomy captures the teamwork concern represented by a requirement, while \fretish{} makes the required behaviour explicit. We report practical lessons and limitations of applying FRET to HRT requirements and identify directions for further work.

\end{abstract}

\section{Introduction}
\label{sec:intro}

The development of critical systems typically begins with a system definition and hazard analysis \cite{mcdermid2002software}, from which requirements are derived to guide subsequent development, verification, and validation~\cite{sommerville1997requirements}. These requirements are commonly expressed in natural language, which is accessible to stakeholders but prone to ambiguity. Formal verification, however, requires precise specifications that can be interpreted by analysis tools~\cite{luckcuck2019formal,azaiez2025revisiting}. Specification is therefore a major bottleneck in the verification pipeline, particularly for autonomous systems~\cite{rozier2016specification}. NASA's Formal Requirements Elicitation Tool (FRET)~\cite{giannakopoulou2020formal} helps bridge this gap by providing the structured natural-language notation \fretish{} and automatically generating temporal-logic semantics~\cite{mavridou2025automated,giannakopoulou2021automated} which can support formal analysis using tools such as NuSMV~\cite{cimatti2002nusmv}, PRISM~\cite{kwiatkowska2002prism}, and CoCoSim~\cite{bourbouh2020cocosim}.


We consider this challenge in the context of Human-Robot Teamwork (HRT), where humans and robots collaborate through different forms of supervision, decision-making, task execution, and information exchange. As HRT is increasingly deployed in applications involving close human--robot interaction, establishing confidence in their safety and functional correctness is essential~\cite{webster2020corroborative}. This creates a need for requirements that are sufficiently precise to support rigorous verification while retaining the teamwork concerns represented in the original natural-language requirements. 


A recent HRT requirements taxonomy~\cite{mavridou2026taxonomy} provides categories for eliciting and analysing such requirements. Taylor et al.~\cite{taylor2026all} applied this taxonomy to natural-language requirements derived from the hazard analysis of a Robot-Assisted Dressing (RAD) system and proposed additional categories arising from the case study. We build on this work by systematically formalising the complete RAD requirement set using FRET and analysing the resulting formal requirements. In particular, we contribute: 


\begin{compactitem}
    \item A corpus of 117 unique \fretish{} requirements formalised from 146 hazard-derived RAD requirements, together with traceability to their source requirements and HRT taxonomy classifications. 
    \item An analysis of the resulting requirements, including decomposition and consolidation, temporal and formal semantics, and relationships across HRT taxonomy categories. 
    \item An initial realisability analysis using FRET, together with practical lessons and limitations identified when applying FRET to HRT requirements. 
\end{compactitem}

\noindent The remainder of the paper is organised as follows. Section~\ref{sec:background} presents background and related work, Section~\ref{sec:methodology} describes the methodology, and  Section~\ref{sec:results} presents the results. Section~\ref{sec:discussion} discusses the findings, and Section~\ref{sec:conclusion} concludes and outlines future work.

\section{Background and Related Work}
\label{sec:background}

\subsection{Robot-Assisted Dressing (RAD)}
\label{subsec:rad}

\begin{figure}[t]
    \centering
    \includegraphics[width=\linewidth]{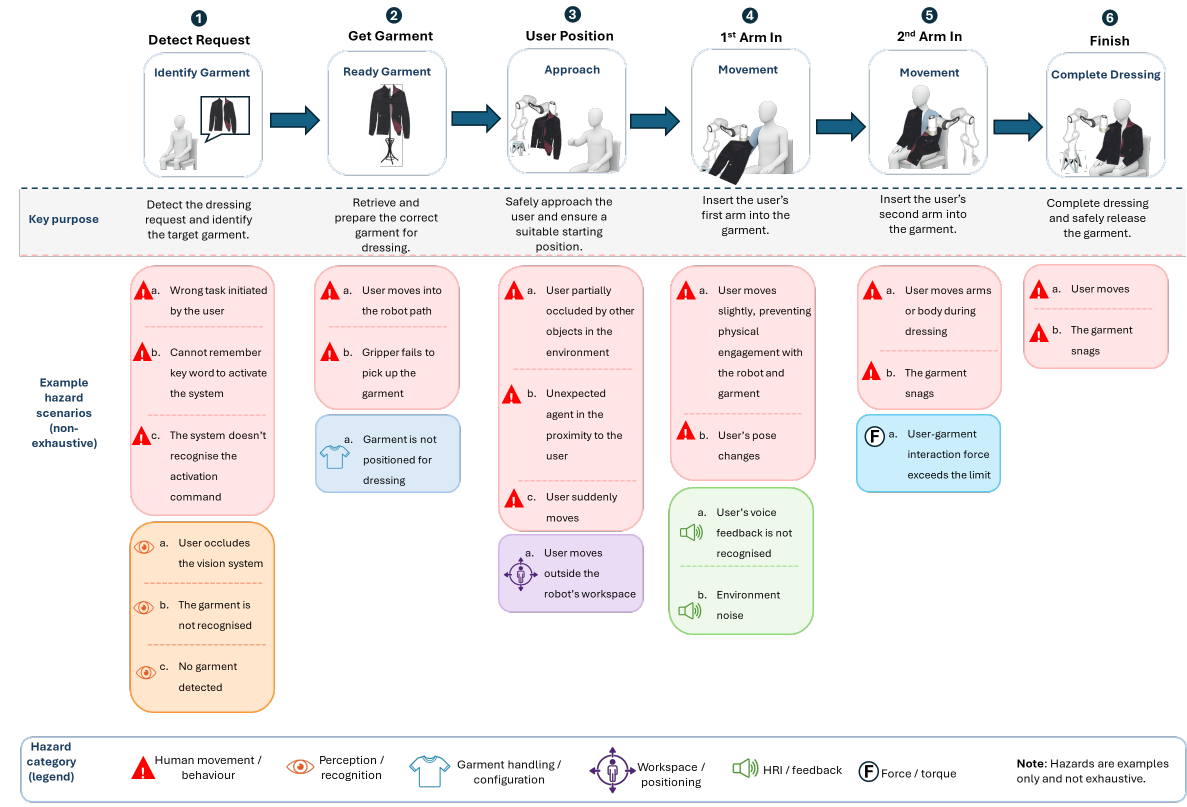}
    \caption{Workflow for the Robot-Assisted Dressing (RAD) use case, including several hazards \cite{taylor2026all}.}
    \label{fig:radworkflow}
\end{figure}

The close physical interaction between the robot and human makes Robot-Assisted Dressing (RAD) both safety-critical and susceptible to execution-time uncertainty. This provides a suitable context for examining how hazards identified through real-world analysis can be translated into formal requirements. The dressing task consists of a sequence of stages, including garment identification, preparation, user approach, arm insertion, and task completion, as illustrated in Fig.~\ref{fig:radworkflow}. Hazards arise throughout these stages from user behaviour, perception and sensing limitations, garment state, workspace constraints, interaction signals, and physical interaction forces~\cite{delgado2021safety}. Taylor et al.~\cite{taylor2026all} used these hazards as the basis for eliciting 146 natural-language requirements, which form the starting point for the formalisation in this paper.


\subsection{Taxonomy of Human-Robot Teamwork Requirements}
\label{subsec:taxonomy}
\begin{figure}[t]
    \centering
    \includegraphics[width=0.9\linewidth]{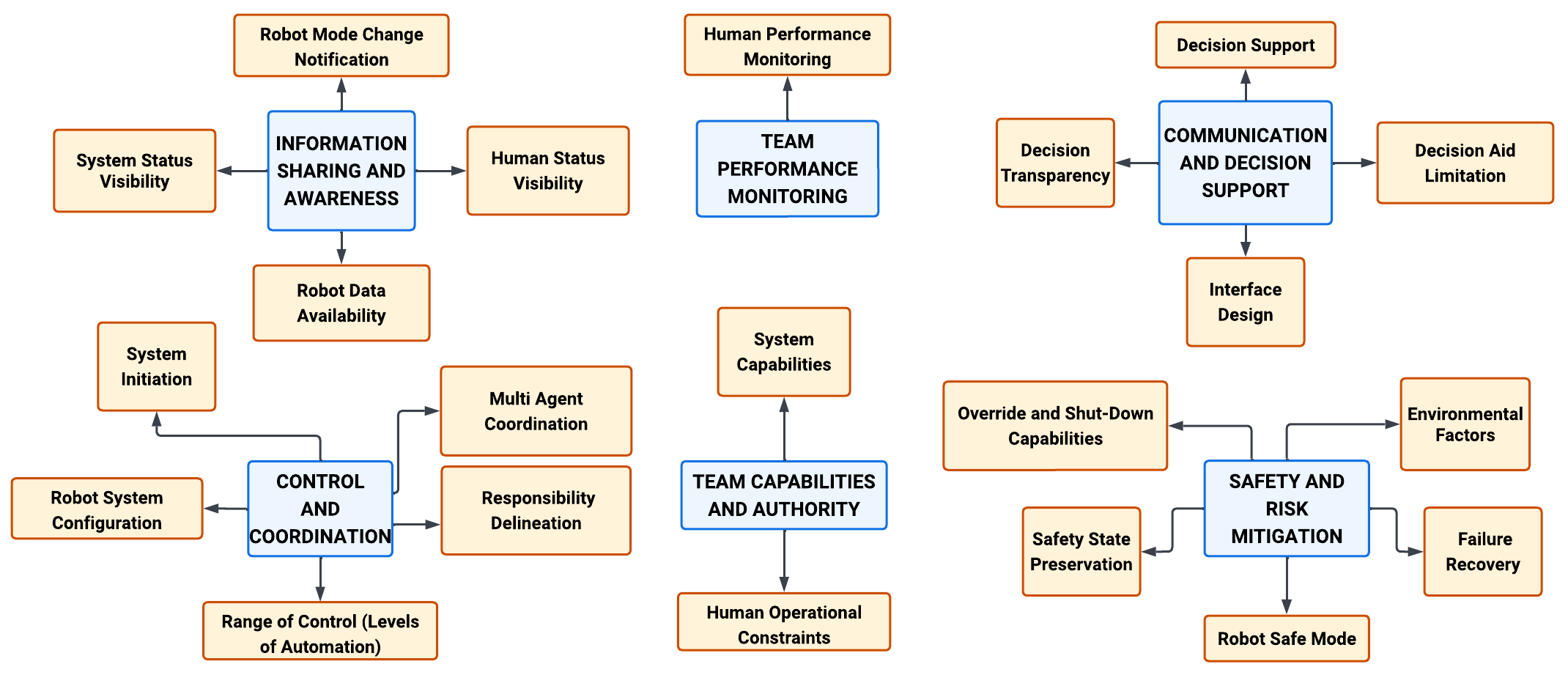}
    \caption{Hierarchy of categories from the Human-Robot Teamwork (HRT) requirements taxonomy in \cite{mavridou2026taxonomy}. The blue boxes represent high level categories and the orange boxes denote low level categories.}
    \label{fig:taxonomyCategories}
\end{figure}

Fig. \ref{fig:taxonomyCategories} outlines the 6 high-level taxonomy categories (in blue) and 21 low-level taxonomy categories (in orange) from \cite{mavridou2026taxonomy}. Our analysis in this paper focuses on the high-level categories only and we summarise them here. We note that \cite{taylor2026all} proposed several new low-level categories but no new high-level ones so our later analysis is in line with the original taxonomy in~\cite{mavridou2026taxonomy}. The high-level categories are:

\begin{compactdesc}
    \item[\category{1. Information Sharing and Awareness:}] This category is concerned with the availability and visibility of system state information, it captures requirements for maintaining situational awareness across the team. Both the human and robot must have access to relevant data, including system status, operational modes, and performance indicators, to support informed participation. The focus is on \textit{passive} awareness: enabling understanding without directly influencing actions. 
 
    \item[\category{2. Control and Coordination}:] This category focuses on the allocation of authority, roles  and responsibility management within the team. Requirements of this kind govern system configuration, task allocation, responsibility delineation, and levels of autonomy. In contrast to \category{Information Sharing and Awareness}, these requirements \textit{actively shape} system behaviour by defining who is responsible for what and when. 
    
    \item[\category{3. Communication and Decision Support:}] \begin{sloppypar} Requirements in this category support human decision-making by structured, interpreted communication. Rather than passively conveying raw state data. They deliver actionable information via decision aids, explanations, alerts, and interface design to support trust, understanding, and timely intervention. \end{sloppypar}
    
    \item[\category{4. Safety and Risk Mitigation:}] Safe operation and hazard management in Human-Robot Teams are addressed via mechanisms such as preventative (e.g. safe workspace design), reactive responses (e.g. safe modes), and recovery-oriented strategies. Requirements in this category constrain system behaviour to prevent harm and manage failures, independent of control allocation. 
    
    \item[\category{5. Team Performance Monitoring:}] This category concerns the continuous assessment of human and robot performance. This includes the monitoring of behaviour, workload, and operational effectiveness, with the main purpose of this category to support adaptive response. Unlike static information sharing, the emphasis here is on \textit{dynamic evaluation} and adjustment of team performance. Detecting degradation and enabling adaptive responses help to maintain safe and effective collaboration. 
    
    \item[\category{6. Team Capabilities and Authority:}] This category captures foundational properties and constraints that bound overall interaction. Specifically, the capabilities of the robotic system and the authority retained by the human. These requirements establish structural limits of teamwork, ensuring that human override capability is preserved while enabling effective collaboration. 
    \end{compactdesc}

\noindent\textbf{RAD and the Taxonomy:} Taylor et al.~\cite{taylor2026all} applied the taxonomy to the 146 hazard-derived RAD requirements and proposed additional low-level categories while retaining the six original high-level categories. \category{Safety and Risk Mitigation} was the largest category, followed by \category{Communication and Decision Support} and \category{Team Capabilities and Authority}. The distribution of source requirements across these categories is reported in Table~\ref{tab:formalisation-coverage}, alongside the corresponding formalisation results.

\subsection{Formal Requirements Elicitation Tool (FRET)}
\label{subsec:fret}
The Formal Requirements Elicitation Tool (FRET) provides a structured natural-language notation, \fretish{}, for formalising requirements \cite{giannakopoulou2020formal}. A \fretish{} requirement consists of the following fields, where an asterisk (*) denotes a compulsory field \cite{mavridou2025automated}:

\vspace{2mm}
\centerline{\fretishComponents}
\vspace{2mm}

The \Scope{} indicates the modes of operation when a requirement should hold. Users can specify the \ConditionF{} under which the requirement should hold, the \Component{} to which the requirement is assigned to, and the \Probability{} associated with the requirement's boolean \ResponseF{} and its associated \Timing{}. The compulsory ``shall'' keyword indicates that the statement is a requirement. Based on the values entered for these specific fields, FRET generates an associated Linear-Time Temporal Logic (LTL) and/or Probabilistic Computation Tree Logic (PCTL*) semantics \cite{giannakopoulou2021automated,mavridou2025automated} (if the \Probability{} field is used). FRET allows users to hierarchically organise requirements using a parent-child relationship but this is purely for organisational purposes and does not define a formal, logical link.

FRET also supports analysis of formalised requirements, including simulation using NuSMV~\cite{cimatti2002nusmv}, generation of verification conditions for tools such as CoCoSim \cite{bourbouh2020cocosim} and R2U2 \cite{aurandt12026resource}, and requirement-level realisability checking~\cite{katis2022capture}. FRET has been applied in safety-critical domains including aerospace~\cite{mavridou2020ten,farrell2022fretting}, robotics~\cite{vazquez2024robotics}, and medical devices~\cite{farrell2024fretting}.

\subsection{Related Work}
\label{subsec:related}

Early work formally verified astronaut--robot teamwork by combining human-agent teamwork models with model checking to analyse collaborative missions~\cite{bordini2009formal}. The Brahms language has similarly been used to provide semi-formal description of HRT in space applications~\cite{ClanceySKH03, Brahms20:AERO}. Vázquez et
al.~\cite{vazquez2024robotics} proposed mission requirement patterns for autonomous mobile robots using FRET, although these do not explicitly capture the human team member. Explainability requirements and patterns have also been studied~\cite{droste2024explanations,refsq} with some formalised using FRET~\cite{eras}. Hendry et al.~\cite{hendry2025roboscene} formalised Human-Robot Interaction (HRI) properties using tock-CSP and RoboScene. 

Complementary work has explored how formal verification can be applied to concrete human-robot collaboration scenarios and the requirements that govern them. Webster et al. proposed a corroborative verification and validation framework that combines model checking, simulation-based testing, and experiments to verify safety and liveness requirements in a collaborative manufacturing handover task \cite{webster2020corroborative}. Lestingi et al. extended formal analysis to service-robot settings by modelling humans, robots, and decision-making processes as hybrid automata and applying statistical model checking to assess mission success and safety properties \cite{lestingi2020formal}. Rafiq et al. combine runtime verification with adaptive decision-making to monitor task execution and dynamically adjust robot behaviour, improving the safety and reliability of human-robot collaboration during assistive dressing tasks \cite{rafiq2025symbolic}. We use the same use case in this paper but systematically formalise the requirements, rather than verifying probabilistic models.

\section{Methodology}
\label{sec:methodology}

We used an iterative methodology to formalise the RAD requirements in \fretish{}. The starting point was the natural-language requirements in~\cite{taylor2026all}, together with the original hazard analysis of the RAD scenario~\cite{delgado2021safety}. The source hazard analysis distinguishes three groups of requirements. Requirements identified by `\texttt{S}' originate from a human-error analysis of hazards associated with specific segments of the dressing task, while those identified by `\texttt{C}' come from a complementary human-error analysis of hazards that may occur across any task segment. Requirements identified by `\texttt{R}' originate from the robot hazard analysis, where potential deviations in robot behaviour are associated with nodes of the system UML model and analysed using the SHARD guide words \emph{Omission}, \emph{Commission}, \emph{Early}, \emph{Late}, and \emph{Value}. These identifiers were retained throughout our formalisation to maintain traceability.

The natural-language requirements were subsequently classified according to the high-level HRT taxonomy categories introduced in Section~\ref{subsec:taxonomy}. We retained both the original hazard-analysis identifier and the HRT taxonomy classification, allowing each formal requirement to be traced both to its source in the hazard analysis and to the HRT taxonomy category that it represents. Where formalisation required a compound source requirement to be decomposed into separate requirements, suffixed identifiers (e.g., \texttt{S17\_a, S17\_b}, and \texttt{S17\_c}) were introduced while retaining the original source identifier.


Before formalisation, a RAD domain expert reviewed all 146 natural-language requirements and provided additional context from the original hazard analysis and dressing scenario. The formalisation was then carried out iteratively by two researchers. Initially, one researcher formalised the \textsc{Safety and Risk Mitigation}  requirements using FRET, which were cross-checked by the second researcher. The remaining taxonomy categories were distributed between the two researchers, who independently formalised their assigned requirements. Throughout this process, ambiguities and formalisation decisions were discussed with the RAD domain expert and a FRET formalisation expert. Every resulting \fretish{} requirement was subsequently cross-checked by both experts: the RAD domain expert assessed consistency with the intended behaviour and operational context, while the FRET expert reviewed the correctness and interpretation of the formalisation. Requirements were revised through further discussion where necessary. Formalisation also required identifying the relevant system states, events, conditions, and responses from the natural-language requirements and representing them as atomic propositions in \fretish{}. These modelling decisions were informed by the original hazard descriptions and RAD domain context, with ambiguities resolved through expert consultation.

The clarification process centred on three issues: (1) whether a natural-language requirement represented a single formal requirement, needed to be decomposed, or whether multiple source requirements expressed the same formal obligation and could be consolidated; (2) which \fretish{} timing construct most accurately represented the intended temporal behaviour; and (3) how ambiguous or underspecified phrases in the natural-language requirements should be interpreted. Where such phrases were identified, their intended meaning was discussed with the RAD domain expert, and the resulting interpretation was reflected in the \fretish{} formalisation. Where natural-language requirements contained qualitative expressions such as ``timely'' or ``high accuracy'', explicit quantitative constraints were introduced based on the RAD domain expert's experience conducting human--robot dressing experiments in the laboratory. For example, ``timely'' was formalised as \textit{within 5 ticks} (R69) or \textit{within 30 seconds} (R94), depending on the operational context, while ``high accuracy'' in S07\_a was represented by a probabilistic threshold of 0.9. Table~\ref{tab:examplereqs} presents a subset of the resulting requirements for illustration. The complete requirements dataset and associated formalisation artefacts are provided in our repository.\footnote{\url{https://github.com/cradlerobotics/RAD}} Across the 10-week formalisation process, approximately 10 hours of discussion with the RAD domain expert were devoted to clarifying ambiguous or underspecified requirements. The number of individual clarification points and their impact on the initial formalisations were not systematically recorded.

\begin{table}[t]
\centering
\scalebox{0.85}{
    \begin{tabular}{|c|p{8.2cm}|p{8.2cm}|}
    \hline
        \textbf{ID} & \textbf{Natural-Language} & \textbf{\fretish{}}  \\ \hline\hline
         \texttt{S\_04a}& The robot shall detect loss of user tracking and immediately stop robot motion & \conditionF{Upon loss\_of\_user\_tracking} \component{Robot} shall \timing{immediately} \responseF{motion\_stopped} \\\hline
         
         \rowcolor{lightgray!60}\texttt{R12} & The robot shall implement 2-stage task initiation verification (using multi-modal sensor input and user queries) to prevent incorrectly confirming a garment has been found &\component{Robot} shall \timing{before item\_identification\_confirmed} \responseF{ internal\_validation\_passed} \\\hline

         \texttt{S22} & If the user requests dressing and the robot detects that they are already wearing the requested garment then the robot shall inform the user that they are already wearing the garment & \conditionF{if dressing_requested \& garment_already_worn} \component{Robot} shall \timing{immediately} \responseF{notify_already_wearing_garment}\\ \hline
        \rowcolor{lightgray!60} \texttt{S05\_c} & The robot shall inform the user that it is in detection mode and remind the user not to move the garment around & \scope{in detectionMode} \component{Robot} shall \timing{at the first timepoint} \responseF{notify_user_of_detection_mode \& remind_user_not_to_move_garment}\\ \hline 

         \texttt{R61} & The robot shall detect when the garment cannot be moved and respond appropriately to avoid prolonged waiting &  \conditionF{Upon (task\_failure\_condition | task\_time\_bound\_exceeded)} \component{Robot} shall \timing{immediately} \responseF{(task\_exception_detected \& task\_paused)} \\
         & & \texttt{+} \conditionF{Upon task\_exception\_detected} \component{Robot} shall \timing{before task\_continued} \responseF{(user\_notified \& appropriate\_user\_input\_requested)}\\\hline

       \rowcolor{lightgray!60}  \texttt{S07\_a} &The robot shall detect potential collisions between the robot and the user with a high level of accuracy &\conditionF{whenever potential\_robot\_user\_collision} \component{Robot} shall \probability{with probability $\geq$ 0.9} \timing{immediately} \responseF{collision\_detected} \\\hline
    \end{tabular}}
    \caption{Selected natural-language requirements with accompanying \fretish{}. We use `\texttt{+}' to denote that a single requirement (e.g. \texttt{R61}) was decomposed into several \fretish{} requirements. }
    \label{tab:examplereqs}
\end{table}

\section{Results and Analysis}
\label{sec:results}

We analyse the formalised requirements from five complementary perspectives. First, we examine the relationship between the original natural-language requirements and the resulting \fretish{} requirements, including cases where requirements were decomposed or consolidated during formalisation using FRET. Second, we examine the temporal constructs that are used in the \fretish{} requirements and how these vary across the high-level HRT taxonomy categories. Third, we examine the formal temporal and probabilistic semantics generated by FRET. Fourth, we use realisability analysis to identify inconsistencies and underspecification within the resulting requirement set. Finally, we examine equivalence and overlap between requirements belonging to different HRT taxonomy categories.


\subsection{Formalisation Coverage and Requirement Transformation}\label{subsec:fretCoverageReqTransformation}

Formalising the hazard-derived natural-language requirements did not always result in a one-to-one mapping between a source requirement and a \fretish{} requirement. In some cases, a single natural-language requirement contained multiple behavioural obligations that needed to be represented separately and was therefore decomposed into multiple \fretish{} requirements. This is a pattern  that has also been observed in related work \cite{vazquez2024robotics}. Conversely, requirements originating from different hazards sometimes expressed the same underlying behavioural obligation and could be represented by a common \fretish{} requirement. We refer to these transformations as \textit{decomposition} and \textit{consolidation}, respectively. 

Table~\ref{tab:formalisation-coverage} contains the formalisation results across the six high-level HRT taxonomy categories. All 146 source requirements are represented, however, this does not result in 146 distinct \fretish{} requirements. Decomposition can cause a single source requirement to give rise to multiple formal requirements, while consolidation occurs when multiple source requirements express the same formal behaviour. After these transformations, the dataset contains 117 unique \fretish{} requirements. \medskip

\begin{table}[t]
\centering
\caption{Formalisation coverage  across the high-level HRT taxonomy categories.}
\label{tab:formalisation-coverage}
\small
\begin{tabular}{|p{6.5cm}|p{2.2cm}|p{2.2cm}|p{2.2cm}|}
\hline
\textbf{Taxonomy Category} &
\textbf{Source \;\;Requirements} &
\textbf{Formalised Requirements} &
\textbf{Unique \fretish{} Requirements} \\
\hline\hline
\rowcolor{lightgray!60} \category{Information Sharing and Awareness}       & 11 & 11 & 9  \\\hline
\category{Control and Coordination}                 & 10 & 10 & 10 \\\hline
\rowcolor{lightgray!60} \category{Communication and Decision Support}       & 42 & 42 & 35 \\\hline
\category{Safety and Risk Mitigation}                & 48 & 48 & 42 \\\hline
\rowcolor{lightgray!60} \category{Team Performance Monitoring}              & 1  & 1  & 1  \\\hline
\category{Team Capabilities and Authority}           & 34 & 34 & 20 \\\hline
\hline
\textbf{Total}                                      & \textbf{146} & \textbf{146} & \textbf{117} \\
\hline
\end{tabular}
\end{table}





\noindent\textbf{\category{Information Sharing and Awareness}} contains 11 source requirements, all of which are represented in \fretish{}, resulting in nine unique formal requirements. The reduction from 11 source requirements to nine unique formal requirements does not indicate that requirements have been removed. Instead, it reflects relationships exposed during formalisation between requirements originating separately in the hazard analysis and the behavioural obligations represented in \fretish{}. \medskip

\noindent\textbf{\category{Control and Coordination}} contains 10 source requirements, all of which are represented in \fretish{}, resulting in 10 unique formal requirements. In contrast to \category{Information Sharing and Awareness}, there is therefore a one-to-one correspondence at the category level between the source requirements and the resulting unique \fretish{} requirements. This indicates that no reduction in the number of requirements through consolidation is observed for this category. \medskip

\noindent\textbf{\category{Communication and Decision Support}} contains 42 source requirements, all of which are represented in the formalised dataset, resulting in 35 unique \fretish{} requirements. The reduction reflects consolidation identified during formalisation, including cases in which requirements originating from separate hazard-analysis records express the same formal behaviour. The working dataset retains these source mappings for traceability even where they contribute to a single unique formal requirement. \medskip

\noindent\textbf{\category{Safety and Risk Mitigation}} is the largest category and it contains 48 source requirements. All of these are represented in \fretish{}, producing 42 unique formal requirements. As with \category{Information Sharing and Awareness}, the mapping is not purely one-to-one. Both decomposition and consolidation occur: some natural-language requirements are split because they contain multiple formal obligations, while other source requirements map to the same formal property. \medskip

\noindent\textbf{\category{Team Performance Monitoring}} contains a single source requirement, which results in one unique \fretish{} requirement. \medskip 

\noindent\textbf{\category{Team Capabilities and Authority}} contains 34 source requirements that are captured as 20 unique \fretish{} requirements. This is the largest proportional reduction between the source and unique formal requirement counts among the six categories. The difference reflects relationships exposed during formalisation, including consolidation of requirements that express common formal behaviour. \medskip
 


An example of decomposition is provided by \texttt{S17} in the \category{Safety and Risk Mitigation} category\footnote{Our repository:
\url{https://github.com/cradlerobotics/RAD/tree/main/data}.}. In the original hazard analysis, \texttt{S17} is associated with a hazard ``User obstructs path'' and recommends that the robot track the user's position, define a boundary at which the user is alerted, and/or provide an emergency-stop procedure~\cite{delgado2021safety}. During formalisation, these behaviours were separated into three requirements, \texttt{S17\_a, S17\_b} and \texttt{S17\_c}, as summarised in Table~\ref{tab:s17-decomposition}.


\begin{table}[t]
\centering
\caption{Decomposition of the source requirement S17 during formalisation. }
\label{tab:s17-decomposition}
\scalebox{0.85}{
\begin{tabular}{|l|p{1.05\linewidth}|}
\hline
\textbf{ID} & \textbf{Natural Language Source Requirement} \\
\hline\hline
\rowcolor{lightgray!60} \texttt{S17} &
Track the user's position, alert the user at a defined boundary,
and/or provide an emergency-stop procedure. \\\hline

\texttt{S17\_a} &
Respond to a potential collision by pausing the robot trajectory
and reminding the user to remain still. \\\hline

\rowcolor{lightgray!60} \texttt{S17\_b} &
Track the user's position and alert the user when they move outside
the robot's operating range. \\\hline

\texttt{S17\_c} &
Provide an emergency-stop response when a collision risk with the
user is detected. \\
\hline
\end{tabular}}
\end{table}


Conversely, consolidation occurs when separately elicited source requirements result in the same \fretish{} requirement. For example, the original hazard analysis defines \texttt{S11} for the situation \emph{``User obstructs EE path''}, with the recommendation \emph{``Robot pauses if user moves when in close proximity''}~\cite{delgado2021safety}. The first obligation derived from \texttt{S17}, \texttt{S17\_a}, similarly captures the required robot response when the user's position creates a potential collision. Although \texttt{S11} and \texttt{S17} originate from different task stages and hazard-analysis entries, formalisation showed that \texttt{S11} and \texttt{S17\_a} express the same requirement:

\noindent{\small \conditionF{When potentialCollisionDetected} \component{Robot} shall \timing{immediately} \responseF{trajectoryPaused \& \\userReminderIssued}}

\texttt{S11} and \texttt{S17\_a} are retained as separate requirement records to preserve traceability, but they contribute only one unique \fretish{} requirement in our analysis. This illustrates how decomposition and consolidation can occur together: formalisation first exposes multiple independently verifiable obligations within a single source requirement, while comparison of the resulting formal requirements can subsequently reveal equivalence to a requirement derived elsewhere in the hazard analysis. When analysing the resulting formal requirement set, these are counted as one unique \fretish{} requirement. 


Overall, these results demonstrate that formalisation is not simply a syntactic translation from natural language into \fretish{}. It can alter the structure of the requirement set by making individual obligations explicit and exposing equivalent requirements. Consequently, the number of natural-language requirements does not equal the number of resulting formal properties. A reduction in the number of unique \fretish{} requirements relative to the source requirement set therefore does not imply loss of requirements, since multiple source requirements may legitimately converge on the same formal property.

\subsection{Temporal Characteristics of the Formalised Requirements}\label{subsect:tempCharacterisitic}

Temporal constraints are an important part of the formalisation. They specify when the response must hold relative to a condition or event. These details are often absent from natural-language requirements and eliciting them usually involves discussion with developers and domain experts as seen in other studies \cite{eras,farrell2022fretting,farrell2024adventures}. We have analysed the timing constructs in the unique \fretish{} requirements identified in Section~\ref{subsec:fretCoverageReqTransformation}. Table~\ref{tab:timing} summarises their distribution across the six high-level HRT taxonomy categories.

\begin{table*}[t]
\centering
\caption{Distribution of \fretish{} \Timing{} constructs across the unique
formalised requirements.}
\label{tab:timing}
\scalebox{0.78}{
\setlength{\tabcolsep}{3pt}
\begin{tabular}{|p{2.4cm}|p{2.45cm}|p{2.55cm}|p{2.9cm}|p{2.15cm}|p{2.5cm}|p{2.3cm}||p{0.9cm}|}
\hline
\Timing{} &
\category{Information Sharing and Awareness} &
\category{Control \;\; \; \;and \;\;\;\;Coordination} &
\category{Communication and \;\;\;\;\;\;\;Decision \;\;\;Support} &
\category{Safety and \;\;\;\;\;\;\;\;\;Risk \;\;\;Mitigation} &
\category{Team \;\;\;Performance Monitoring} &
\category{Team \;\;\;Capabilities and \;\;\;Authority} &
\textbf{Total} \\\hline\hline
\rowcolor{lightgray!60} \timing{immediately}           & 6 & 5 & 12 & 16 & 0 & 5 & \textbf{44} \\\hline
\timing{before}                & 1 & 4 & 12 & 14 & 0 & 7 & \textbf{38} \\\hline
\rowcolor{lightgray!60} \timing{always}                & 1 & 1 & 4  & 6  & 1 & 2 & \textbf{15} \\\hline
\timing{within}                & 0 & 0 & 1  & 4  & 0 & 5 & \textbf{10} \\\hline
\rowcolor{lightgray!60} \timing{eventually}            & 0 & 0 & 1  & 1  & 0 & 1 & \textbf{3} \\\hline
\timing{Unspecified}                 & 0 & 0 & 4  & 0  & 0 & 0 & \textbf{4} \\\hline
\rowcolor{lightgray!60} \timing{after}                 & 0 & 0 & 1  & 0  & 0 & 0 & \textbf{1} \\\hline
 \timing{never}                 & 0 & 0 & 0  & 1  & 0 & 0 & \textbf{1} \\\hline
\rowcolor{lightgray!60} \timing{at the first timepoint}& 1 & 0 & 0  & 0  & 0 & 0 & \textbf{1} \\\hline
\hline
\textbf{Total}                 & \textbf{9} & \textbf{10} & \textbf{35} & \textbf{42} & \textbf{1} & \textbf{20} & \textbf{117} \\
\hline
\end{tabular}}
\end{table*}

The \category{Safety and Risk Mitigation} category contains 42 unique \fretish{} requirements and exhibits a broad range of timing constructs. The most frequent are \timing{immediately} (16 requirements ) and \timing{before} (14 requirements), together accounting for 71.4\% of the category.  

This distribution reflects two distinct forms of temporal constraint within the safety requirements. The frequent use of \timing{immediately} captures reactive behaviour in which the robot must respond directly to a detected event. For example, loss of user tracking requires robot motion to stop immediately (\texttt{S\_04a} in Table \ref{tab:examplereqs}). In contrast, \timing{before} expresses ordering constraints in which a condition must be satisfied prior to another event, such as validating an item before its identification is confirmed (e.g. \texttt{R12} in Table \ref{tab:examplereqs}). The bounded \timing{within} requirements introduce explicit time limits on responses. 

\category{Information Sharing and Awareness} contains nine unique \fretish{} requirements. Six use  \timing{immediately}, while \timing{before}, \timing{always}, and \timing{at the first timepoint} each occur once. The prevalence of \timing{immediately} reflects requirements in which changes in robot or user state need to be made available to another member of the human--robot team without an intervening delay. This emphasises the urgency that one would expect with conditions related to \category{Information Sharing and Awareness}.


\category{Control and Coordination} contains 10 unique \fretish{} requirements and is concentrated around two timing constructs. Five requirements use \timing{immediately}, while four use \timing{before}; the remaining requirement uses \timing{always}. Thus, nine of the 10 requirements express either an immediate response or an ordering constraint. Compared with \category{Safety and Risk Mitigation}, this category uses a narrower range of timing constructs, despite sharing the prominence of \timing{immediately} and \timing{before}.

\category{Communication and Decision Support} contains 35 unique \fretish{} requirements. The two most common are \timing{immediately} and \timing{before}, each occurring in 12 requirements (34.3\%). Four unique \fretish{} requirements do not specify an explicit timing construct; in these cases, FRET's generated semantics requires the response to hold \timing{eventually} within the applicable scope. The prominence of both \timing{immediately} and \timing{before} reflects the combination of responsive communication and ordering constraints in this category: some information must be communicated as soon as a relevant condition occurs, while other actions must take place before a subsequent decision or interaction. 

\category{Team Capabilities and Authority} contains 20 unique \fretish{} requirements. Seven use \timing{before}, five use \timing{immediately}, and five use the bounded \timing{within} construct. Two requirements use \timing{always} and one uses \timing{eventually}. The comparatively frequent use of \timing{within} distinguishes this category from most of the others, indicating several requirements in which a capability or response must be provided within an explicit temporal bound. The single requirement in \category{Team Performance Monitoring} uses \timing{always}; given the size of this category, no broader timing pattern can be inferred from it. 


Across the complete formalised requirement set, \timing{immediately} and \timing{before} are the dominant constructs, occurring in 44 (37.6\%) and 38 (32.5\%) of the 117 unique \fretish{} requirements, respectively. Together they account for 82 requirements (70.1\%). The timing profile therefore differs from that reported in previous FRET case studies. Farrell et al.~\cite{farrell2024adventures} found that the most commonly used timing options across four case studies were \timing{eventually}, \timing{always}, and \timing{at the next timepoint}. In contrast, the RAD requirements make substantial use of \timing{immediately} and \timing{before} across several HRT taxonomy categories. The RAD requirements also use bounded \timing{within} constraints, which occur in 10 requirements.

These findings have implications beyond the distribution of FRETISH timing constructs. The prominence of \timing{immediately} and \timing{before} highlights two important classes of obligations in the RAD scenario: reactive responses to changes in human or robot state, and ordering constraints that govern when actions may proceed. Making these distinctions explicit helps identify properties for subsequent verification, particularly whether safety responses occur when required and whether prerequisite conditions are satisfied before task progression. The temporal profile also illustrates how formalisation exposes behavioural constraints that may be implicit in natural-language requirements. While these observations are specific to RAD, the analysis provides a basis for examining whether similar patterns arise in other HRT applications.


\subsection{Formal Semantics of the \fretish{} Requirements}\label{subsec:semanticsFRET}

Formalisation in \fretish{} provides a structured representation of each requirement from which FRET automatically generates formal temporal-logic semantics. Table~\ref{tab:formal-semantics} illustrates this transformation using four examples. \texttt{S04\_a} demonstrates an immediate response, \texttt{R77} an ordering constraint, \texttt{R103} a bounded response, and \texttt{S07\_a} an explicitly probabilistic response. The first three requirements are \category{Safety and Risk Mitigation}, while \texttt{S07\_a} is classified under \category{Team Capabilities and Authority}.


\begin{table*}[t]
\centering
\caption{Examples of the transformation from natural-language requirements to \fretish{} and the corresponding infinite-trace LTL and PCTL* semantics generated by FRET*. }
\label{tab:formal-semantics}
\scalebox{0.8}{
\begin{tabular}{|p{0.16\textwidth}|p{0.97\textwidth}|}
\hline
\textbf{ID and\;\;\;\;\;\;\; Representation} & \textbf{Requirement} \\
\hline\hline

\rowcolor{lightgray!60} \texttt{S04\_a}: &
Upon loss of user tracking, the robot shall stop its motion. \\

\rowcolor{lightgray!60} \textbf{\fretish{}:} &
\conditionF{Upon loss\_of\_user\_tracking}
\component{Robot} shall
\timing{immediately}
\responseF{motion\_stopped} \\

\rowcolor{lightgray!60} \textbf{LTL:} &
\texttt{\detokenize{((G (((! loss_of_user_tracking) & (X loss_of_user_tracking)) ->}} \;\;\;\;\;\;\;\;\;\;\;\;\;\;\;\;\;\;\;\;\;\;\;\;\;\;\;\;\;\;\texttt{\detokenize{(X motion_stopped))) &}} \texttt{\detokenize{(loss_of_user_tracking -> motion_stopped))}} \\
\hline

\texttt{R77}: &
The robot shall ensure that the dressing task is only completed when
it has been performed correctly. \\

\textbf{\fretish{}:} &
\component{Robot} shall
\timing{before dressing\_complete}
\responseF{correct\_dressing\_performed} \\

\textbf{LTL:} &
\texttt{\detokenize{(correct_dressing_performed V (! dressing_complete))}} \\
\hline

\rowcolor{lightgray!60} \texttt{R103}: &
The robot shall perform corrective action in a timely manner when
corrective action is required. \\

\rowcolor{lightgray!60} \textbf{\fretish{}:} &
\conditionF{Upon corrective\_action\_required}
\component{Robot} shall
\timing{within 30 seconds}
\responseF{corrective\_action\_applied} \\

\rowcolor{lightgray!60} \textbf{LTL:} &
\texttt{\detokenize{((G (((! corrective_action_required) &}} \texttt{\detokenize{(X corrective_action_required)) ->}} \texttt{\detokenize{(X (F[0,30] corrective_action_applied)))) &}} \texttt{\detokenize{(corrective_action_required ->}} \texttt{\detokenize{(F[0,30] corrective_action_applied)))}}  \\

\hline
\texttt{S07\_a}:
&
The robot shall detect potential collisions between the robot and the
user with a high level of accuracy. \\[1mm]
\textbf{\fretish{}:} & 
\conditionF{Whenever potential\_robot\_user\_collision}
\component{Robot} shall
\probability{with probability $\geq 0.9$}
\timing{immediately}
\responseF{collision\_detected}\\ 

\textbf{PCTL*:}
& 
\texttt{\detokenize{P>=1[(G (potential_robot_user_collision =>}}
\texttt{\detokenize{(P>=0.9[collision_detected])))]]}}
\\
\hline
\end{tabular}}
\end{table*}


\texttt{S04\_a} expresses that robot motion stops upon loss of user tracking using the \timing{immediately} timing. The generated LTL captures both the case where loss of tracking holds initially and transitions in which it subsequently becomes true. Here, \timing{immediately} requires the response to hold at the same logical timepoint as the triggering event, rather than at the next timepoint. This interpretation concerns the discrete-time semantics of the requirement and does not, by itself, establish a zero-latency response in the physical RAD system. \texttt{R77} uses \timing{before} to require correct dressing to have occurred before the dressing task can be considered complete. \texttt{R103} illustrates a bounded temporal requirement: the natural-language notion of corrective action being performed in a timely manner is made explicit in \fretish{} as \timing{within 30 seconds}, and this bound is preserved in the generated LTL semantics. 

\texttt{S07\_a} illustrates the use of an explicit probability bound. The natural-language requirement specifies that potential robot--user collisions should be detected with a ``high level of accuracy''. During formalisation, this qualitative expression was made explicit as a probability of at least 0.9. Unlike the preceding examples, the resulting probability bound constrains the likelihood of satisfying the required response, and FRET therefore generates a PCTL* property that preserves this bound. 


10 unique \fretish{} requirements contain probabilistic features. Nine occur in \category{Communication and Decision Support}, while one, \texttt{S07\_a}, occurs in \category{Team Capabilities and Authority}. 
Temporal and probabilistic relationships that are implicit or qualitative in the source requirements must become explicit in \fretish{}. 
This is apparent in \texttt{R103}, where ``timely'' is replaced by a concrete 30-second bound, and in \texttt{S07\_a}, where a ``high level of accuracy'' is represented by an explicit probability threshold. 

\subsection{Realisability Analysis}\label{subsec:realisability}

Even with formalisation tools like FRET, there can be inconsistencies and errors introduced. To address this, FRET supports realisability checking of requirements \cite{katis2022capture} to detect conflicts. A requirement set is unrealisable when no implementation can satisfy all requirements simultaneously due to conflicts or excessive constraints. Realisability analysis helps to detect specification errors, and FRET's approach decomposes the analysis into smaller sub-problems to improve scalability while preserving soundness \cite{mavridou2026requirements}. In this way, it decomposes the requirement set into ``connected components'' (CC) or groupings of related requirements. We conducted realisability checking on a per-category basis. FRET uses Kind2, an SMT-based model checking and verification tool, as the underlying engine for realisability analysis. 

In \category{Safety and Risk Mitigation}, Kind2 terminated unexpectedly on one CC, timed out on two others and identified four CCs causing unrealisability where configurations are violated in the initial state (i.e. at the first time point). This indicates missing assumptions in the requirement set that cause underspecification.  The \category{Team Capabilities and Authority} category experienced several timeouts during realisability checking on various CCs and three errors in realisability that relate to underspecification in the initial state.  The \category{Control and Coordination} category also had several realisability errors related to underspecification in the initial state, as did \category{Communication and Decision Support}. Since this kind of realisability issue arose across categories we believe that there is significant underspecification for the initial configuration of the RAD system. Understanding this limitation and improving the requirements set to robustly capture the initial configuration is future work. 

In \category{Communication and Decision Support} the realisability checker identified a conflict between  \texttt{C02\_a} and \texttt{S02}, and \texttt{C02\_a} and \texttt{S03\_b} as in Table \ref{tab:conflict}. Both conflicts arise because \texttt{confirm\_command} becomes true when either \texttt{ambiguity\_present} or \texttt{risk\_present} hold according to \texttt{C02\_a}. This doesn't have the added restriction to satisfy \texttt{use\_multiple\_modalities}, as in \texttt{S02} and \texttt{S03\_b} before confirming the command so there is a conflict. Table~\ref{tab:conflict} presents a proposed revision, \texttt{C02\_a\_fix}, which requires multiple modalities to be used before confirming the command. Realisability checking was repeated following this modification to assess the revised requirements. However, the outcome of this subsequent check is not reported here, and we therefore cannot establish whether the modification resolved all conflicts. Further changes to individual requirements may also introduce conflicts elsewhere in the requirement set, highlighting the importance of iterative realisability checking. 
\begin{table}[t]
    \centering
    \scalebox{0.85}{
    \begin{tabular}{|c|p{\textwidth}|}
    \hline
         \textbf{ID} & \textbf{\fretish{}}  \\ \hline\hline
        \rowcolor{lightgray!60}  \texttt{C02\_a} & \conditionF{whenever ambiguity\_present | risk\_present} \component{Robot} shall \timing{immediately} \responseF{confirm\_command} \\ \hline
         \texttt{S02} & \conditionF{whenever speech\_commands\_incomplete} \component{Robot} shall \probability{with probability > 0.9} \timing{before confirm\_command} \responseF{use\_multiple\_modalities} \\\hline
        \rowcolor{lightgray!60}  \texttt{S03\_b} & \conditionF{whenever background\_noise\_overpowering} \component{Robot} shall \probability{with probability > 0.9} \timing{{before confirm\_command} \responseF{use\_multiple\_modalities}}\\ \hline\hline
         \texttt{C02\_a\_fix} & \conditionF{whenever ambiguity\_present | risk\_present} \component{Robot} shall \timing{before confirm\_command} \responseF{use\_multiple\_modalities}\\\hline
    \end{tabular}}
    \caption{Conflicting requirements. We note that the \Probability{} field of \texttt{S02} and \texttt{S03\_b} is not used during realisability checking since it only checks the LTL representation.}
    \label{tab:conflict}
\end{table}

There were no realisability issues detected within \category{Information Sharing and Awareness} or  \category{Team Performance Monitoring}. In fact the latter had only a single requirement to check. Realisability checking is not yet supported for probabilistic requirements, for these it removes the probability when doing realisability and only considers the LTL semantics for the requirement.

\subsection{Equivalence and Overlap Between Requirements}\label{subsec:equivalence}

Formalisation exposed relationships between requirements assigned to different high-level HRT taxonomy categories. These relationships were not limited to exact formal equivalence. Requirements could share identical or closely related \fretish{} expressions while representing different concerns in the originating hazard analysis, or could arise from a common hazard while imposing distinct behavioural obligations. Table~\ref{tab:cross-category} illustrates four forms of cross-category relationship identified in the dataset.

\begin{table*}[t]
\centering
\caption{Example relationships between requirements from
different HRT taxonomy categories.}
\label{tab:cross-category}
\scalebox{0.8}{
\setlength{\tabcolsep}{4pt}
\begin{tabular}{|p{0.15\textwidth}|p{0.35\textwidth}|p{0.17\textwidth}|p{0.47\textwidth}|}
\hline
\textbf{Requirements} &
\textbf{Categories} &
\textbf{Relationship} &
\textbf{Formalisation outcome} \\
\hline\hline

\rowcolor{lightgray!60} \texttt{C05\_a}, \texttt{C05\_b} &
\category{Team Capabilities and Authority}, \;\;\;\;\;\;\;\;\;\;\;\;\;
\category{Communication and Decision Support} &
 Identical \fretish{}; source semantics differ &
Both map to the same \fretish{} expression concerning
\texttt{user\_speech\_detected\_during\_audio\_feedback} and
\texttt{audio\_feedback\_stopped}, although the source requirements
distinguish listening for user speech from stopping audio feedback. \\

\hline

\texttt{S16\_b}, \texttt{R103} &
\category{Team Capabilities and Authority},
\category{Safety and Risk Mitigation} &
Conceptual overlap; formally distinct &
Both concern adaptation or corrective behaviour, but impose different
obligations: S16\_b concerns trajectory adaptation following user
movement, whereas R103 requires corrective action within a bounded
30-second interval. \\

\hline

\rowcolor{lightgray!60} \texttt{S19}, \texttt{S23}, \texttt{S27}, \texttt{S28},  \texttt{S32_a}, \texttt{S32\_b} &
\category{Team Capabilities and Authority}, \;\;\;\;\;\;\;\;\;\;\;\;\;
\category{Communication and Decision Support} &
Repeated cross-category decomposition &
A common user-pose-change context gives rise to complementary
requirements: the \_a requirements concern robot adaptation to the
user's position, while the \_b requirements concern guidance to the
human team member. \\

\hline

\texttt{C02\_a}, \texttt{C02\_b}, \texttt{C02\_c} &
\category{Communication and Decision Support}, 
\category{Team Capabilities and Authority}, 
\category{Safety and Risk Mitigation} &
Common provenance; distinct obligations &
Requirements derived from the same source hazard are separated into
distinct obligations concerning command confirmation, validation before
execution, and prevention of unsafe command execution. \\

\hline
\end{tabular}}
\end{table*}

Across the complete formalised dataset, three unique \fretish{} properties occur in more than one HRT taxonomy category, as determined by exact matching of the \fretish{} expressions. 
For example, \texttt{C05\_a} and \texttt{C05\_b} have identical \fretish{} expressions, but the source requirements capture different concerns: \texttt{C05\_a} requires the robot to listen for user speech while providing audio feedback, whereas \texttt{C05\_b} requires the robot to stop its audio feedback when the user speaks. The shared formalisation therefore captures the response required once speech is detected, but in its current form does not preserve the source-level distinction between actively listening for speech and responding to its detection.  

Other cross-category relationships do not produce identical formal properties. \texttt{S16\_b} and \texttt{R103} overlap conceptually because both concern adaptation of robot behaviour in response to changing conditions, but they remain formally distinct. \texttt{S16\_b} specifies trajectory adaptation in response to user movement, whereas \texttt{R103} places a bounded temporal requirement on corrective action. Similarly, the repeated decomposition of \texttt{S19, S23, S27, S28}, and \texttt{S32} separates a common user-pose-change situation into complementary obligations: robot adaptation is classified under \category{Team Capabilities and Authority}, while guidance to the user is classified under \category{Communication and Decision Support}.  The \texttt{C02} requirements illustrate a further relationship between taxonomy classification and formalisation. \texttt{C02\_a, C02\_b}, and \texttt{C02\_c} originate from the same source hazard but are assigned to different HRT taxonomy categories because they address different aspects of the interaction. 

These results show that HRT taxonomy classification and formalisation capture complementary aspects of a requirement. The taxonomy identifies the HRT concern to which a requirement relates, whereas \fretish{} captures the behavioural obligation imposed on the system. Consequently, requirements from different taxonomy categories may converge on the same formal expression, while requirements arising from the same source context may require different formal properties. Exact syntactic agreement in \fretish{} should therefore be interpreted together with the originating natural-language requirement and its hazard context before requirements are treated as semantically equivalent or consolidated.

\section{Discussion}
\label{sec:discussion}
In this section, we discuss how we decomposed and consolidated requirements, and  how FRET helped us to identify and reduce ambiguity. We also explore links between the HRT taxonomy and the requirements formalisation. Finally, we reflect on lessons learned and any limitations that we encountered.



\subsection{Decomposition, Consolidation and Reducing Ambiguity}\label{subsec:formalisationRequirementRefinement}


Hazard-derived requirements may combine several behavioural obligations, while requirements originating from different hazards may express equivalent behaviour. Decomposition makes these individual obligations explicit, whereas consolidation identifies common behaviour across the requirement set.  These transformations  create an important traceability consideration. For example, decomposing \texttt{S17} allows its individual obligations to be analysed separately, while the equivalence between \texttt{S11} and \texttt{S17\_a} shows that distinct hazard contexts can lead to the same formal property. Consequently, consolidation should not remove the source mappings: a single formal property may retain links to multiple requirements and hazards that explain why that behaviour is required.

Formalisation exposed temporal assumptions that were implicit or underspecified in the natural-language requirements. \fretish{} requires explicit temporal specifications, and the discussion in Section~\ref{subsect:tempCharacterisitic} shows that timing is an important aspect of RAD requirements. 
Bounded timing requirements make the clarification problem particularly visible. For example, \texttt{R103} originally requires corrective action in a ``timely manner'' which our formalisation process replaced by an explicit 30-second bound. FRET exposes such ambiguity but it cannot automatically determine what the correct bound should be. This remains a domain and requirements-engineering decision. 

This highlights a broader lesson: 
a syntactically valid formal requirement is not necessarily a correct requirement. Formalisation is valuable because it makes assumptions precise enough to inspect and challenge, but domain expertise remains necessary to establish whether those assumptions capture the intended HRT behaviour. Formalisation 
exposes behavioural structure and relationships that may be less apparent in the natural-language dataset, while preserving the hazard context through traceability. 

\subsection{Taxonomy and Formal Behaviour}\label{subsec:taxonomyFormalBehaviour}

The cross-category relationships identified in Section~\ref{subsec:equivalence} show that HRT taxonomy classification and formalisation capture different aspects of a requirement. The taxonomy describes the HRT concern represented by a requirement, whereas \fretish{} specifies the behavioural obligation imposed on the system. 
This distinction is particularly apparent where identical \fretish{} expressions occur across categories. The \texttt{C05\_a/C05\_b} example shows that syntactic equivalence does not necessarily imply equivalence of the source requirements: information about whether the robot must listen for user speech or respond once speech is detected can be lost at the chosen level of formalisation. Conversely, the repeated decomposition of user-pose-change requirements shows how a common interaction context can give rise to complementary obligations concerning robot adaptation and communication with the human. 

These observations suggest that taxonomy classification provides useful context for interpreting formal requirements rather than an alternative representation of them. 
More generally, combining HRT-oriented classification with formalisation provides complementary views of teamwork concern represented by a requirement and the behaviour that the system is required to exhibit.

\subsection{Practical Lessons in Formalising HRT Requirements and Limitations}\label{subsec:lessonsLimitations}

Our experience also identified practical limitations in formalising HRT requirements. FRET is effective at exposing ambiguity and forcing temporal assumptions to become explicit, but determining the intended behaviour  depends on domain-specific knowledge. For example, terms such as ``timely'' require an externally agreed bound, while requirements containing several behaviours may need to be decomposed before they can be represented clearly. A syntactically valid \fretish{} requirement cannot be treated as evidence that the intended semantics have been confirmed. The diagrammatic and temporal logic semantics provided in FRET can help the user to assess whether the formalisation does indeed match their intended semantics.


We had several instances of requirements comprising parts of other requirements. For example, \texttt{R42} is contained within \texttt{R88}. Directly consolidating these requirements would not maintain traceability from the hazard analysis so it would be more convenient if we could simply ``call'' \texttt{R42} within \texttt{R88}. This would also simplify the task of maintaining these requirements since a change to an expression in \texttt{R42} could also necessitate a change to the same expression in \texttt{R88}. The Mu-FRET extension to FRET that supports the refactoring of requirements does enable this kind of functionality \cite{luckcuck2026mu} and exploring its use throughout this case study is left as future work. Similar refactoring features could have helped with the repeated formalisation of the two-stage failure recovery behaviour defined in \texttt{R11}, \texttt{R14}, \texttt{R16}, \texttt{R19} and \texttt{R61} to reduce the need for repeated editing as ambiguities were clarified. Similar inconveniences were found during the formalisation of an aircraft engine software controller \cite{farrell2022fretting} which motivated the development of Mu-FRET \cite{luckcuck2026mu}.

Several requirements referred to specific variables being \texttt{valid}, \texttt{accurate}, \texttt{correct}, and/or \texttt{safe}. During expert discussions, it became clear that each of these terms had slightly different properties but some depended on each other. For example, in order to be \texttt{correct}, a variable should also be \texttt{valid} and \texttt{accurate}. These kinds of relationships cannot currently be specified using FRET but an extension of this kind would be useful and allow for more robust checks on the requirement set to be carried out.

Finally, our findings are based on a single RAD case study and on an iterative formalisation process involving domain-expert clarification. The observed timing distributions, decomposition and consolidation patterns, and cross-category relationships cannot be generalised to HRT requirements more broadly. Nevertheless, the study shows the value of combining structured HRT classification, traceability to the originating hazards, and formalisation: together they expose ambiguities and modelling decisions that may remain hidden when the natural-language requirements are considered alone.



\section{Conclusion and Future Work}\label{sec:conclusion}

This paper investigated the formalisation of hazard-derived requirements for the Robot-Assisted Dressing case study using FRET, together with their classification using an existing HRT requirements taxonomy. Through an iterative process involving domain-expert clarification, 146  source requirements were represented by 117 unique \fretish{} requirements. The analysis showed that formalisation was not a one-to-one translation: requirements were decomposed where they contained multiple behavioural obligations and consolidated where different source requirements expressed the same formal behaviour. Formalisation also made temporal and probabilistic assumptions explicit and enabled further analysis of the resulting requirements. 


Our findings distinguish between general challenges of requirements formalisation and observations specific to the RAD human-robot teamwork context. More generally, formalisation exposed ambiguities, required explicit temporal and probabilistic assumptions, and revealed opportunities for decomposition and consolidation while preserving traceability to the original requirements. For HRT, the analysis highlighted the importance of reactive safety responses, task-ordering constraints, and requirements governing interaction between human and robot team members. The HRT taxonomy and \fretish{} provided complementary perspectives: the taxonomy identified the teamwork concern, while \fretish{} made the associated behavioural obligations explicit. Cross-category relationships further demonstrated that identical formal expressions do not necessarily preserve the distinctions present in their originating HRT requirements.

Future work will apply the approach to further HRT case studies to examine whether the patterns observed for RAD generalise across different forms of human--robot teamwork. We also plan to investigate how the generated formal properties and their traceability to hazards and HRT concerns can support subsequent assurance activities, including verification and validation of human--robot team behaviour.

\section*{Acknowledgements}
This work was supported by the Research Institute on Verified Trustworthy Software Systems (VeTSS), the Royal Academy of Engineering, and the Centre for Robotic Autonomy in Demanding and Long Lasting Environments (CRADLE) under EPSRC grant EP/X02489X/1. 

Gunjari Chakraborty was also supported by a Royal Academy of Engineering and Google DeepMind ``Research Ready'' summer internship. 

\bibliographystyle{eptcs}
\bibliography{bibliography}
\end{document}